# Polarization effects in the nonlinear interference of down-converted photons

ANNA PATEROVA[1,2], HONGZHI YANG[1], CHENGWU AN[3], DMITRY KALASHNIKOV[1], AND LEONID KRIVITSKY[1,*]

[1] *Institute of Material Science and Engineering, Agency for Science Technology and Research (A\*STAR), 138634 Singapore*
[2] *School of Electrical and Electronic Engineering, Nanyang Technological University, 639798 Singapore*
[3] *National Metrology Centre, Agency for Science Technology and Research (A\*STAR), 118221 Singapore*
*\*Leonid_Krivitskiy@imre.a-star.edu.sg*

**Abstract:** We study polarization effects in the nonlinear interference of photons generated via frequency non-degenerate spontaneous parametric down conversion. Signal and idler photons generated in the visible and infrared (IR) range, are split in different arms of a nonlinear Michelson interferometer. The interference pattern for signal photons is detected, and it is shown to be dependent on the polarization rotation of idler photons, introduced by a birefringent sample. Based on this concept, we realize two new methods for measurement of sample retardation in the IR range by using well-developed and inexpensive components for visible light. The accuracy of the methods meets current industry standards. The developed IR polarimetry technique is relevant to material research, optical inspection, and quality control.

## 1. Introduction

Interference effects in spontaneous parametric down-conversion (SPDC) have been extensively studied over the last few decades. In addition to the fundamental interest, revealing counterintuitive features of quantum mechanics, they also find practical applications in quantum communication [1], computation [2] and metrology [3-7]. In this work we consider the effect of the nonlinear interference of down-converted photons, also known as "induced coherence" [8, 9]. When signal and idler photons generated in two nonlinear crystals are superimposed, the interference is observed in the intensity and coincidence counts [9]. In contrast to the classical case, the interference pattern for signal/idler photons depends on phases and amplitudes of all the three interacting photons: the signal, the idler, and the pump [10]. This effect is particularly useful in metrology applications when the sample properties are to be measured in the detection challenging spectral range, for instance in the infrared (IR). When the unknown sample is inserted in the path of idler photons, one can infer its properties in the IR range from the interference pattern for signal photons in the visible range. This concept was recently implemented in several practical applications, including IR imaging [12], IR spectroscopy [13-17] and IR tunable optical coherence tomography (OCT) [18].

Earlier works studied influence of various experimental factors on the nonlinear interference, including the effects of spatial and temporal overlap of SPDC modes [19-21], linear losses and dispersion [8, 9, 22] introduced in the interferometer. Analysis of relevant polarization effects, however, is less extensive [23-25]. Grayson et al. demonstrated that signal photons acquire the non-local Pancharatnam phase, which was introduced into the path of idler photons by a set of retardation elements [24]. Recently, Lahiri et al. used the nonlinear Mach-Zehnder interferometer and introduced a polarizer into the path of signal photons and an attenuator into the path of idler photons to study the degree of polarization in such a system [25].

In this work, we perform a systematic analysis of polarization effects in the nonlinear interference of down-converted photons. We derive explicit relations between polarization transformations of idler photons and the interference pattern of signal photons. Based on this principle, we propose and experimentally realize two new methods for characterization of sample retardation in the IR range using well-developed components for visible light.

## 2. Theory

We consider the nonlinear Michelson interferometer, shown in Figure 1a [17, 18]. SPDC photons (signal and idler) generated at the first pass of the pump through a nonlinear crystal enter the interferometer. The pump, signal and idler photons travel in different paths after being separated by corresponding dichroic mirrors (DM$_1$, DM$_2$). The photons are reflected by the mirrors and recombined at the crystal. The reflected pump generates another pair of down-converted photons, which interferes with the pair traveled in the interferometer.

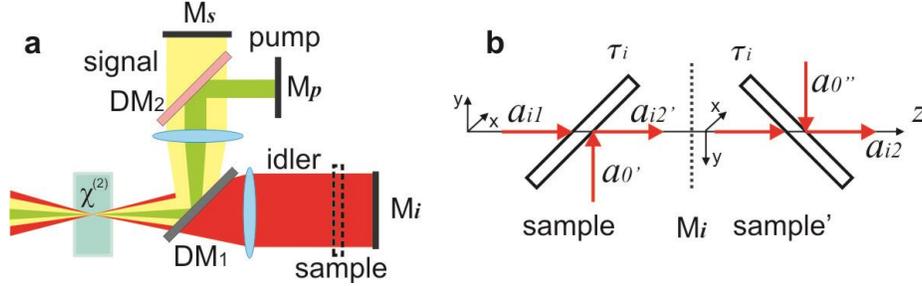

Fig. 1. (a) The nonlinear Michelson interferometer. Pump beam (green) generates SPDC photons (yellow and red), which are separated into different arms by the dichroic mirror DM$_1$. Dichroic mirror DM$_2$ separates signal and pump photons. Mirrors M$_s$, M$_p$ and M$_i$ reflect all the photons back into the crystal, where the pump generates another pair of photons. Interference of signal photons is then detected. (b) The beam splitter model which accounts for the double pass through the sample and reflection from the mirror M$_i$; $\tau_i$ is the amplitude transmission of the beam splitter. Mode $a_{i1}$ transforms into mode $a_{i2}$ by injecting vacuum modes $a_0$ and $a_{0''}$ from open ports of the beam splitter. Mirror M$_i$ inverts the Cartesian coordinate system (x-y).

The state vector of SPDC photons generated in the single crystal can be written as [26]:

$$|\psi\rangle = |vac\rangle + \eta \sum_{k_s, k_i} \sum_{\sigma_s, \sigma_i} F(\vec{k}_s, \vec{k}_i) a^+_{k_s, \sigma_s} a^+_{k_i, \sigma_i} |vac\rangle \qquad (1)$$

where $\eta \ll 1$ is the SPDC conversion factor; $a^+_{k_s,\sigma_s}$, $a^+_{k_i,\sigma_i}$ are photon creation operators for signal and idler modes with wave vectors $\vec{k}_s$ and $\vec{k}_i$, polarizations $\sigma_s$ and $\sigma_i$, respectively; $F(\vec{k}_s, \vec{k}_i)$ is the two-photon amplitude, which defines spatial and spectral properties of SPDC photons; sub-indices $s$ and $i$ indicate signal and idler modes, respectively [27, 28].

Let us consider type-0 phasematching when pump, signal and idler photons have the same linear polarization $\sigma_s = \sigma_i = \sigma$ (the also theory applies to type-I and type-II phasematching). According to Eq. (1), for the single spatial and temporal mode, the state vector of SPDC photons created at the first $|\psi\rangle_1$ and at the second $|\psi\rangle_2$ passes of the pump through the crystal is given by the superposition state [8, 9, 17, 18]:

$$|\Psi\rangle = |\psi\rangle_1 + |\psi\rangle_2 = |vac\rangle + \eta a^+_{s1,\sigma} a^+_{i1,\sigma} |vac\rangle + \eta e^{i\varphi_p} a^+_{s2,\sigma} a^+_{i2,\sigma} |vac\rangle, \qquad (2)$$

where $\varphi_p$ is the phase acquired by the pump in the interferometer [29, 30].

When the sample is inserted into the path of idler photons, the state vector in Eq. (2) changes. According to the beam splitter model [31], and assuming that idler modes *i1* and *i2* are matched, the photon annihilation operator for the idler mode *i2* is given by:

$$a_{i2,\sigma} = e^{i\varphi_i}\left(\tau_i^2 a_{i1,\sigma} + \tau_i\sqrt{1-\tau_i^2}\,a_{0',\sigma} + \sqrt{1-\tau_i^2}\,a_{0'',\sigma}\right), \quad (3)$$

where $a_{0',\sigma}$ and $a_{0'',\sigma}$ denote vacuum fields entering from open ports of the beam splitters, $\tau_i$ is the amplitude transmission coefficient of the beam splitter, $\varphi_i$ is the acquired phase. The count rate for signal photons is given by $P_s \propto \langle\Psi|E_s^{(-)}E_s^{(+)}|\Psi\rangle$. Following calculations, described in details in [17, 18], we obtain the following expression:

$$P_s \propto 2\left[1+|\tau_i|^2|\mu(\Delta t)|\cos\left(\varphi_p - \varphi_s - \varphi_i + \arg\tau_i^2 + \arg\mu(\Delta t)\right)\right], \quad (4a)$$

$$\mu(\Delta t) = \int|F(\Omega)|^2 e^{-i\Omega(\Delta t)}d\Omega, \quad (4b)$$

where $\mu(\Delta t)$ is the normalized correlation function of the SPDC, and $\Delta t$ is the time delay between propagation times of signal and idler photons in the interferometer, $\Omega$ is the frequency detuning [32]. The count rate $P_s$ for the signal photons depends on the transmission coefficient of idler photons, and on the phases of the signal, the idler, and the pump photons.

Let us now redefine the factor $|\tau_i|^2$, taking into account the polarization properties of the sample. Without the loss of generality, we consider the sample to be a generic retardation waveplate. We use the Jones matrices formalism [33, 34] and introduce the corresponding transformation matrix $T_{wp} = \begin{pmatrix} \tau_m e^{i\delta/2} & 0 \\ 0 & \tau_m e^{-i\delta/2} \end{pmatrix}$, where $\delta$ is the retardation between extraordinary and ordinary waves, and $\tau_m$ is the transmission coefficient of the sample (it accounts for reflection, absorption and scattering in the sample). Then, the rotation of the Cartesian coordinate system before and after the sample is given by: $J = R(\theta)T_{wp}R(-\theta)$, where $R(\theta) = \begin{pmatrix} \cos\theta & -\sin\theta \\ \sin\theta & \cos\theta \end{pmatrix}$ is the coordinate rotation matrix, and $\theta$ is the orientation of the optical axis of the sample. Figure 1b shows the detailed description of the propagation of the idler mode *i1*. It accounts for a double pass of the photons through the sample and reflection by the mirror $M_i$. The Jones matrix for this system is given by $J' = R(\theta)T_{wp}MT_{wp}R(\theta) = R(\theta)T_{wp}T_{wp}R(-\theta)$, where $M$ is the transformation matrix of the mirror $M_i$. We can then re-write the resulting Jones matrix in the following form:

$$J' = \tau_m^2\begin{pmatrix} t & r \\ -r^* & t^* \end{pmatrix}, \quad (5)$$

where $t$ and $r$ are the complex amplitude transmission and reflection coefficients of the sample, respectively:

$$t = \cos\delta + i\sin\delta\cos 2\theta$$
$$r = i\sin\delta\sin 2\theta \quad (6)$$

We assume that horizontally polarized (along *x*-axis) idler photons are created in the first and the second pass of the pump through the nonlinear crystal. The idler photon in mode *i1* has an initial polarization vector $\vec{e} = \begin{pmatrix} 1 \\ 0 \end{pmatrix}$ [33, 34]. After propagation through the sample, the polarization vector is modified as follows: $J'\begin{pmatrix}1\\0\end{pmatrix}=\tau_m^2\begin{pmatrix}t\\-r^*\end{pmatrix}$. The modulus of the amplitude transmission function for the horizontally polarized component is given by:

$$|\tau_i|^2 = |\tau_m|^2 |t| = |\tau_m|^2 \sqrt{\cos^2\delta + \sin^2\delta \cos^2 2\theta}, \tag{7}$$

Since the interference can only be observed for the horizontally polarized component of the idler photon, we can substitute the transmission function $|\tau_i|^2$ in Eq. (4a) by Eq. (7). Thus we obtain the following expression for the count rate of signal photons:

$$P_s \propto 2\left[1 + |\tau_m|^2 |t| |\mu(\Delta t)| \cos\left(\varphi_p - \varphi_s - \varphi_i + \arg\tau_m^2 + \arg\mu(\Delta t) + \arg t\right)\right] \tag{8}$$

Let us analyze this expression in more details and show how it can be used for measurement of the sample retardation.

*Method 1. Relative phase shift of interference fringes.* We set the fast optical axis of the sample parallel to the initial horizontal polarization of idler photons ($\theta=0°$). Next, we rotate the sample at 90 degrees, so that the slow axis becomes parallel to the polarization of idler photons ($\theta=90°$). In these two cases, the idler photon experiences refractive indices $n_o$ and $n_e$, respectively (we assume $n_o<n_e$). According to Eq. (8), the relative phase shift between interference fringes of signal photons is proportional to the sample retardation $\delta$:

$$\delta = \varphi_i(\theta=90°) - \varphi_i(\theta=0°) = \frac{2\pi(n_e - n_o)}{\lambda_i} l_m, \tag{9}$$

where $l_m$ length of the sample accounting for a double pass of idler photons, and $\lambda_i$ is the wavelength of idler photons.

*Method 2. Visibility of interference fringes.* From Eq. (8) we obtain the following expression for the visibility (contrast) of the nonlinear interference:

$$V = |\tau_m|^2 |\mu(\Delta t)||t| = |\tau_m|^2 |\mu(\Delta t)| \sqrt{\cos^2\delta + \sin^2\delta \cos^2 2\theta}. \tag{10}$$

From Eq. (10) it follows that the maximum and minimum values of the visibility are observed at $\theta=0°/90°$ and $\theta=45°$, respectively. Then, the retardation can be directly found from the ratio of minimum and maximum values of the visibility:

$$\frac{V_{\min}}{V_{\max}} = \frac{\sqrt{\cos^2\delta}}{\sqrt{\cos^2\delta + \sin^2\delta}} = |\cos\delta|. \tag{11}$$

Note that similar relation is valid when the sample is inserted into the path of signal photons (see Supplementary materials).

The two methods allow measuring the optical retardation of the sample at the wavelength of idler photons from the measurements of the interference of signal photons. Note that Eqs. (9), (11) account for the double pass of idler photons through the sample, and $\delta/2$ gives sample retardation in the single pass. Next, we describe the experimental realization of the two methods and discuss the obtained results.

### 3. Experiment

The experimental setup is shown in Figure 2. A continuous wave (cw) laser with 532 nm wavelength is used as a pump. The laser is focused by a lens $F_1$ ($f = 200$ mm) into a periodically poled Lithium Niobate (PPLN) crystal, where SPDC occurs. The phase matching conditions are set to generate the signal photons at $\lambda_s=809.2$ nm and the idler photons at $\lambda_i=1553$ nm ($\Lambda_{PPLN}=7.4$ µm poling period; temperature $T_{PPLN}=403$ K). Signal and idler beams are separated by a dichroic mirror $DM_2$ into visible and IR arms, respectively, and collimated by lenses F' (both $f = 75$ mm). Then pump and signal beams are split by the dichroic mirror $DM_3$ into separate channels. All the three beams are reflected into the crystal by mirrors $M_{s,p}$ (silver coated) and $M_i$ (gold coated). The reflected pump beam generates another pair of SPDC photons. In the detection part, the pump and idler photons are filtered out by the dichroic mirror ($DM_1$), the notch filter (NF) and the bandpass filter (BP, 809.2 ±0.6 nm). Signal beam is collimated by lens $F_2$ ($f$=200 mm) and detected by an avalanche photodiode (APD) or a CCD camera preceded by lenses $F_3$ ($f$=50 mm) and $F_4$ ($f$=100 mm), respectively. The CCD camera is used to facilitate the setup alignment.

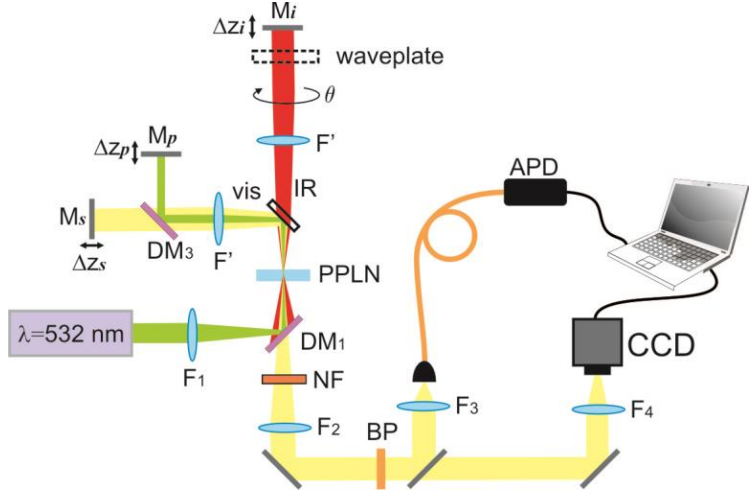

Fig. 2. The experimental setup. The cw-laser pumps the PPLN crystal, where SPDC occurs. The PPLN is set to generate signal and idler photons in the visible and IR range, respectively. The photons are split by the dichroic mirror $DM_2$ into different arms. Pump and signal photons are separated by the dichroic mirror $DM_3$. All the photons are reflected by the mirrors $M_s$, $M_p$ and $M_i$. Filters $DM_1$, NF, and BP filter the detected signal photons. The interference is detected either by the avalanche photodiode (APD) or by the CCD camera. Mirror $M_s$ is mounted on the translation stage for adjustment of the optical path $\Delta z_s$. The sample is inserted into the path of idler/ signal photons. Mirrors $M_i$ and $M_p$ are placed on piezo-translators for fine scans of interference fringes.

The interference of the signal photons is observed once the interferometer arms are equalized within the coherence length of the SPDC (see Supplementary materials). The mirror $M_s$, mounted on a translation stages (step size ~1 µm), is used to equalize the interferometer arms ($\Delta z_s$). Test samples are mounted into a rotation stage and inserted into the interferometer arms. Fine scans of the interference of signal photons are performed by translating the mirror $M_i$ or $M_p$ ($\Delta z_{i,p}$), mounted onto the piezo stages (step size ~ 2 nm). The interference fringes are measured at different orientations of the sample $\theta$. The interferometer was not actively stabilized and was observed to be stable during the measurements.

## 4.  Results and discussion

### 4.1 Sample in the path of idler photons

First, we test our method with samples of known retardation, namely quarter- (QWP) and half- (HWP) waveplates designed for 1550 nm. For the initial calibration we measure the interference fringes without a sample in the IR arm (see Supplementary materials). Then, the sample is inserted into the IR arm with its optical axis set parallel to the polarization of the idler photons ($\theta$=0°). The change of the optical path length is compensated by translation of the mirror Ms. Once the optimal position of the mirror is found, we perform fine scans by translating the mirror Mi in the idler channel using a piezo-stage. Measurements are taken at different orientations of the optical axis of the sample $\theta$.

Figs. 3a,b show interference fringes of signal photons measured for the orientation of the sample at $\theta=0°$, $45°$ and $90°$. The relative phase shift between patterns at $\theta=0°$ and $90°$ is equal to $\delta$. As expected from Eq. (9), the shift of the interference pattern is $\lambda_{i/2}$ and $\lambda_i$ for the QWP and HWP, respectively (idler photons travel twice through the sample). This demonstrates the first method for measurement of the sample retardation. Summary of the obtained results is given in Table I.

Figure 3c shows the dependence of the visibility of the interference on the orientation of the sample $\theta$. In accordance with Eq. (10) the interference fringes have maximum visibility at $\theta=0°$ and $90°$, while at $\theta=45°$ the visibility reaches its minimum. For the QWP at $\theta=45°$ the visibility is nearly equal to zero ($V=0.005\pm0.005$). Solid lines in Figure 3c correspond to theoretical curves for waveplates designed for 1550 nm (sample absorption data is shown in Table SI of Supplementary materials). Thus we realize the second method for measurement of the sample retardation. The measurement results are summarized in Table I.

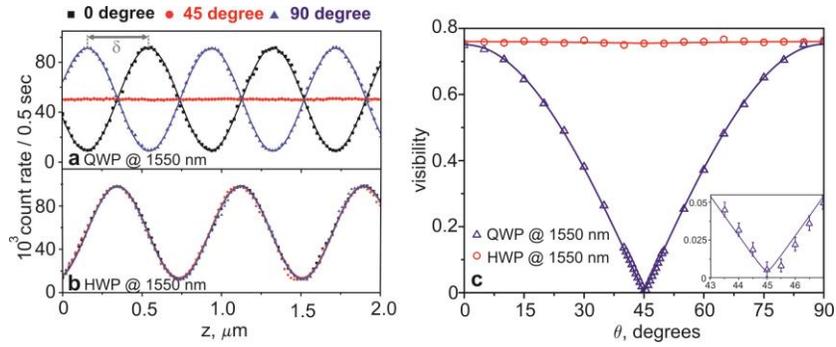

Fig. 3. The count rate of signal photons at $\lambda_s=809.2$ nm versus translation of the mirror $M_i$ in the idler channel for (a) QWP at 1550 nm and (b) HWP at 1550 nm. The orientation of the optical axis at $0°$ (black squares), $45°$ (red dots) and $90°$ (blue triangles). Points are experimental data, and solid lines are fits by Eq. (8) ($R^2>0.99$). The relative phase shift between interference patterns at $\theta=0°$ and $90°$ is equal to retardation $\delta$. **(c)** The dependence of the visibility on the sample orientation $\theta$ for QWP (blue triangles) and HWP (red circles) at 1550 nm. Points are experimental data, and solid lines are fits by Eq. (10) ($R^2>0.99$). The inset shows zoomed results for QWP at 1550 nm at visibility values close to zero.

Next, we perform measurements of samples with arbitrary retardations. We used HWP and QWP designed for operation at 532 nm. These waveplates operate as unconventional retarders for the probing beam at $\lambda_i=1553$ nm. Figs. 4a,b show interference fringes of signal photons measured for the two samples at orientations $\theta=0°$, $45°$, and $90°$. Corresponding retardation values are calculated by Eq. 9, and the results are shown in Table I. Figure 4c shows the dependence of the visibility on the orientation of the two samples. Solid lines are fits with Eq.(10) (sample absorption data shown in Table SI of Supplementary materials). The retardation is inferred from the ratio of the minimum and maximum values of the visibility, see Eq. (11). The measurement results are summarized in Table I.

**Table I Results of retardation measurements at 1553 nm by the two methods (idler mirror scan).**

| Sample | Retardation at $\lambda_i$=1553 nm $\delta/2$ (single path), $\pi$ | | |
|---|---|---|---|
| | Phase shift method | Visibility method | Reference value* |
| HWP at 1550 nm | $1.004 \pm 0.006$ | $0.980 \pm 0.030$ | $1.000 \pm 0.006$ |
| QWP at 1550 nm | $0.495 \pm 0.006$ | $0.498 \pm 0.002$ | $0.500 \pm 0.006$ |
| HWP at 532 nm | $0.322 \pm 0.006$ | $0.321 \pm 0.006$ | Not available |
| QWP at 532 nm | $0.172 \pm 0.006$ | $0.173 \pm 0.010$ | Not available |

* data provided by the manufacturer.

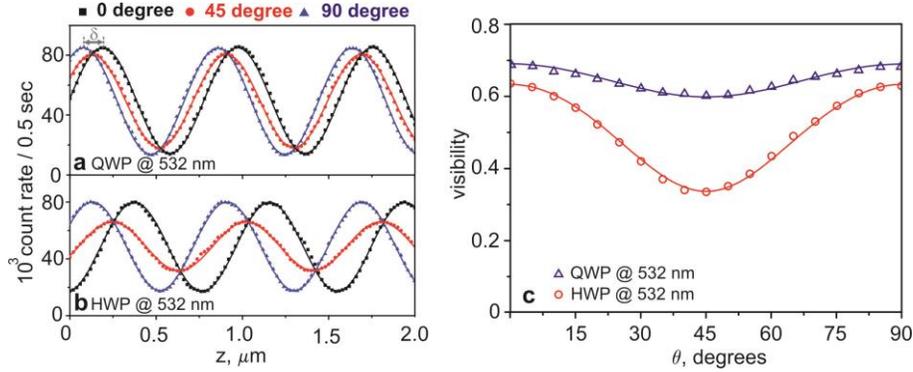

Fig. 4. The count rate of signal photons at $\lambda_s$=809.2 nm versus translation of the mirror $M_i$ in the idler channel for (a) QWP and (b) HWP at 532 nm inserted in the path of idler photons. The orientation of the optical axis of the sample is at 0° (black squares), 45° (red dots) and 90° (blue triangles). Points are experimental data, and solid lines are fits by Eq. (8) ($R^2$>0.99). The relative phase shift between interference patterns at $\theta$=0° and 90° is equal to $\delta$. (c) The dependence of the visibility on the orientation of the sample $\theta$ for QWP (blue triangles) and HWP (red circles) designed for 532 nm. The solid curves show fits by Eq. (10) ($R^2$=0.98).

Next, we investigate the dependence of interference fringes on the translation of the mirror in the pump arm. Figure 5 shows interference fringes obtained for QWP and HWP designed for operation at 1550 nm and 532 nm for different orientation of the optical axis (at $\theta$=0°, 45°, and 90°). We perform fine scans of the mirror $M_p$, placed on the piezo-stage. The pump wavelength $\lambda_p$ now defines the periodicity of the pattern. Similar to the above procedure, we infer the sample retardation from the phase shift and the visibility ratio. The results are summarized in Table II. They agree with the ones reported in Table I.

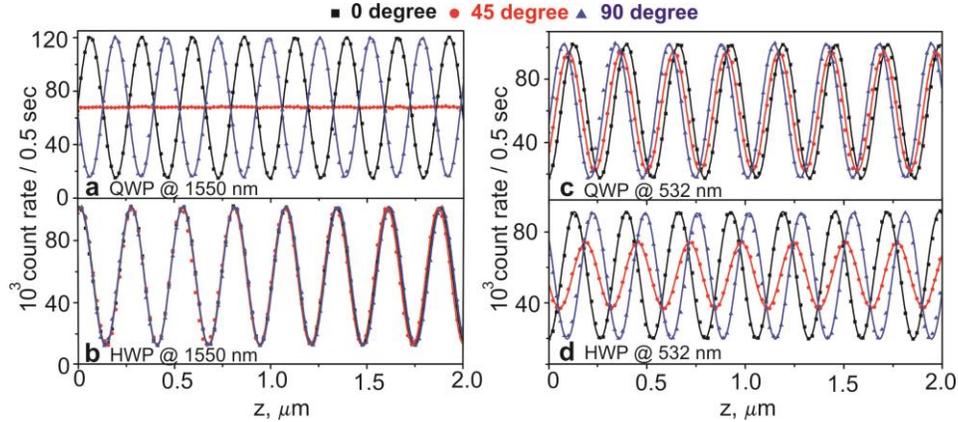

Fig. 5 The count rate measured by translating pump mirror $M_p$ for (a) QWP at 1550 nm, (b) HWP at 1550 nm, (c) QWP at 532 nm, (d) HWP at 532 nm inserted in the idler arm with orientations of the optical axis at 0° (black), 45° (red) and 90° (blue). Points are experimental data, and solid lines are fits by Eq. (8) ($R^2$>0.99).

**Table II Results of retardation measurements at 1553 nm by the two methods (pump mirror scan).**

| Sample | Retardation at $\lambda_i$=1553 nm $\delta/2$ (single path), $\pi$ | | |
|---|---|---|---|
|  | Phase shift method | Visibility method | Reference value* |
| HWP at 1550 nm | 0.994 ± 0.006 | 0.983 ± 0.010 | 1.000 ± 0.006 |
| QWP at 1550 nm | 0.507 ± 0.006 | 0.499 ± 0.001 | 0.500 ± 0.006 |
| HWP at 532 nm | 0.325 ± 0.006 | 0.323 ± 0.006 | Not available |
| QWP at 532 nm | 0.175 ± 0.006 | 0.162 ± 0.012 | Not available |

\* data provided by the manufacturer.

## 4.2 Sample in the path of signal photons

Next, we study polarization transformations by the sample, when it is placed in the signal channel. We use the zero-order QWP designed for 800 nm and perform scans of the mirror in the idler channel. Similarly to previous measurements, we observe the phase shift of the interference fringes at different orientations of the sample $\theta$, see Figure 6a. Also, we measure the visibility dependence at each orientation of the sample (Method 2), see Figure 6b. The retardation is then inferred from the shift of the interference fringes (Method 1) and from the ratio of minimum/ maximum visibilities (Method 2) The results are summarized in Table III.

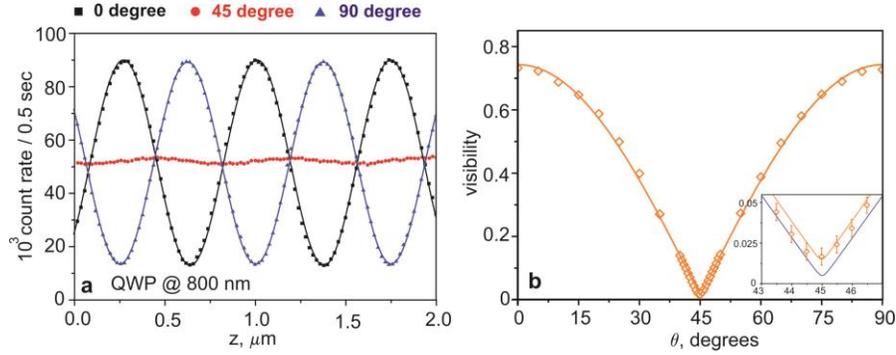

Fig. 6. (a) The count rate measured by translating the mirror $M_i$ in the idler channel when the QWP at 800 nm is inserted in the signal channel. Orientations of the optical axis are at 0° (black squares), 45° (red dots) and 90° (blue triangles). (b) The dependence of the visibility of the interference on the orientation of the sample. Points are experimental data, and solid lines are fits with Eq. (8) in (a) and Eq. (11) in (b) (both $R^2 > 0.99$). The inset shows zoomed visibility values near zero for QWP at 800 nm (yellow circles and line), and for QWP at 1550 nm (blue line) for reference.

Table III Results of retardation measurements at 800 nm by the two methods (idler mirror scan).

| Sample | Retardation at $\lambda_s$=800 nm $\delta/2$ (single path), $\pi$ | | |
|---|---|---|---|
|  | visibility method | phase shift method | Reference value |
| QWP at 800 nm | 0.485 ± 0.001 | 0.491 ± 0.006 | 0.500 ± 0.006 |

## 4.3 Discussion

Based on the obtained experimental data we are now able to compare the two methods for characterization of samples retardation. The "phase shift" method (Method 1) is fast, as we infer the retardation from two sets of fringes at different orientations of the sample. However, the interferometer should be well stabilized during the measurements to minimize drift of the fringes due to external fluctuations (thermal and mechanical). The best experimental accuracy in determination of the retardation $\delta$ is about ±0.006$\pi$.

In contrast, the "visibility method" (Method 2) is much more tolerable to the phase drifts, as they do not strongly affect the visibility. However, the measurement takes longer time than the "phase shift" method. The accuracy of the method is defined by the contrast between the minimum and maximum visibilities achieved in the experiment. In this method, the best achieved experimental accuracy is at the order of ±0.002$\pi$, which also accounts for the accuracy in setting up the sample orientation. Both the methods provide the accuracy comparable to the conventional methods reported by the manufacturers.

As our results show, the period of the interference fringes is defined by scanning configuration: it is given either by $\lambda_p$ or by $\lambda_i$, depending on which mirror is scanned. The configuration with the scan of the mirror in the pump beam may represent a certain advantage when the translation range of the piezo-stage is limited. We also note that when the sample is inserted in the path of the signal photon, the retardation can be measured at the visible wavelength within the same configuration.

## 5. Conclusions

In conclusion, we performed a detailed study of polarization effects in the nonlinear interferometer. We showed that the change of the polarization of idler photons affects both the visibility and phase of the interference fringes of signal photons. This effect allows us inferring the sample retardation at the frequency of idler photons from (a) shift of the interference fringes and (b) from the visibility ratio. The suggested "phase shift" and the "visibility" methods allow characterization of the sample retardation with the accuracy up to $\Delta\delta=\pm0.002\pi$, which meets the requirements for the optical industry. This technique can be readily extended to IR polarimetry of samples with optical activity, such as Faraday rotators and chiral media.

Further extension of the operating wavelengths of the presented methods can be easily achieved by choice of the periodic poling of the crystals and/ or by tuning crystal temperature. Operation within the range of 1,5-4,3 um was shown in our earlier works with LiNb crystals [15-18].

Also, active control over polarization of probing photons allows enhancing the signal-to-noise ratio by compensating polarization changes in the realistic samples. This idea forms the basis for polarization-sensitive optical coherence tomography (PS-OCT) [35] and polarization-sensitive quantum optical coherence tomography (PS-QOCT) [36], which have already been set forth as high-contrast methods for the birefringence measurement of layered samples. Consequently, our technique can be used for the PS-OCT development, extending the conventional methods to the mid-and far-IR range.

### Funding

This work was supported by the Quantum Technology for Engineering program (QTE) of A*STAR, project # A1685b0005; A.P. acknowledges support from the Singapore International Graduate Award (SINGA) fellowship.

### Acknowledgments

We acknowledge fruitful discussions with Sergei Kulik, Rainer Dumke, and Berthold-Georg Englert.

## Appendix. Supplementary materials

*A.1 Visibility of the interference*

The state vector of type-0 SPDC the in Eq. (2) is given by:

$$|\Psi\rangle = |\psi\rangle_1 + |\psi\rangle_2 = |vac\rangle + e^{i(\varphi_s+\varphi_i)}|H\rangle_{s1}|H\rangle_{i1} + e^{i\varphi_p}|H\rangle_{s2}|H\rangle_{i2}, \quad (A1)$$

Where $|H\rangle$ is the single-photon Fock state with horizontal polarization. When the waveplate is introduced into the path of signal photons the state vector in Eq. (A1) is given by:

$$|\Psi\rangle = |vac\rangle + e^{i(\varphi_s+\varphi_i)}J'|H\rangle_{s1}|H\rangle_{i1} + e^{i\varphi_p}|H\rangle_{s2}|H\rangle_{i2}, \quad (A2)$$

Then, the polarization state vector of the signal photons $|H\rangle_{S1}$ is transformed as $J'\begin{pmatrix}1\\0\end{pmatrix}_{s1} = \begin{pmatrix}t_s\\-r_s^*\end{pmatrix}$, where index $s$ denote signal photons. Then, the state vector of the SPDC photons in Eq. (A2) is given by:

$$|\Psi\rangle = e^{i(\varphi_s+\varphi_i)}\left(t_s|H\rangle_{s1} - r_s^*|V\rangle_{s1}\right)|H\rangle_{i1} + e^{i\varphi_p}|H\rangle_{s2}|H\rangle_{i2}. \quad (A3)$$

Assuming that idler photon modes are aligned $i1=i2=i$, the count rate of the signal photons is given by:

$$P_s \propto \langle\Psi|a_s^+ a_s|\Psi\rangle = \left|e^{i(\varphi_s+\varphi_i)}t_s + e^{i\varphi_p}\right|^2 + \left|e^{i(\varphi_s+\varphi_i)}r_s^*\right|^2 = \\ = 2 + 2|t_s|\cos(\varphi_s+\varphi_i-\varphi_p+\arg t_s) \quad (A4)$$

In Eq. (A4) the interference visibility is given by $V=|t_s|$ as which is similar to Eq. (10).

*A2. Alignment of the interferometer*

Initially, the interferometer is balanced, and the interference pattern is observed around the zero position of the translation stage of $M_s$. The reference interference pattern is shown by black squares in Figure A1. Once the sample is inserted in the path of idler photons, the mirror $M_s$ has to be moved to compensate for the introduced optical delay. Figure A1 shows interference pattern after introducing samples into the path of idler photons.

For the HWP at 1550 nm, QWP at 1550 nm, HWP at 532 nm phase shifts are approximately the same, as they are all purchased from the same supplier (Thorlabs) and have close values of the optical thicknesses. The QWP at 532 nm is purchased from another vendor (DayOptics) and the optical thickness is somewhat smaller.

Figure A2 shows the shift of the reference interference pattern due to the introduction of the QWP at 800 nm into the path of signal and idler photons. As the waveplate is designed for 800 nm its losses for signal photons are smaller than for idler.

By measuring the interference at the position of maximum visibility of the interference pattern it is possible to define the transmission function $|\tau_m|^2$, see Table AI. This data is used to plot theoretical curves for the visibility function in Figures 3, 4, and 6. The experimental data in Table AI is in a good agreement with reference measurements, obtained with a conventional IR spectrophotometer (Shimadzu UV3600). The accuracy of the transmission coefficient measurement is ±0.012, which is higher than shown in [17] due to nearly 5-times increase in the visibility of the interference pattern.

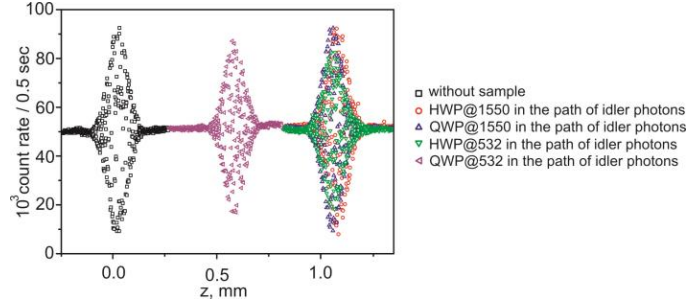

***Fig. A1*** *Shift of the interference pattern after the introduction of the sample in the path of idler photons. Data for HWP at 1550 nm, QWP at 1550 nm, HWP at 532 nm and QWP at 532 nm with the orientation of the optical axis at θ=0°.*

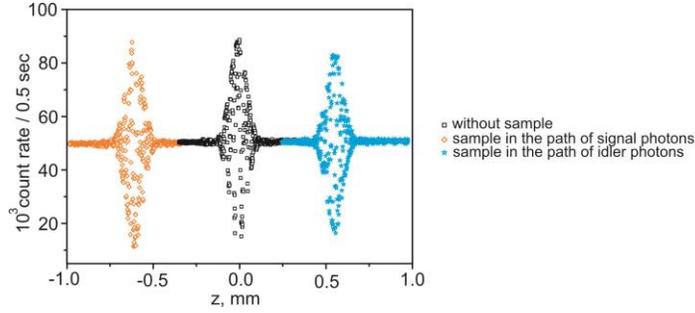

***Fig. A2*** *Shift of the interference pattern due to the change of the optical path for the signal and idler photons after insertion of the HWP at 800 nm with the orientation of the optical axis at θ=0°.*

**Table AI Transmission coefficient $|\tau_m|^2$ for different samples.**

| Sample | Probe wavelength, nm | Transmission coefficient $|\tau_m|^2$ | |
|---|---|---|---|
| | | Measured | Reference* |
| HWP at 1550 nm | 1553 | 0.985±0.012 | 0.998 |
| QWP at 1550 nm | 1553 | 0.980±0.012 | 0.995 |
| HWP at 532 nm | 1553 | 0.857±0.018 | 0.848 |
| QWP at 532 nm | 1553 | 0.903±0.016 | 0.891 |
| QWP at 800 nm | 809.2 | 0.986±0.013 | 0.998 |

\* obtained with an IR spectrophotometer (Shimadzu UV3600)